\documentclass[10pt,a4paper,onecolumn]{vupreprint}
\usepackage[bindingoffset=0.5cm,textheight=25cm,hdivide={2.5cm,*,3cm}, vdivide={*,24cm,*}]{geometry}
\usepackage{amsmath}
\usepackage{amsfonts}
\usepackage{amssymb,mathbbol}
 \usepackage[numbers,square,comma,sort&compress]{natbib}
 \usepackage{fancyhdr}
 \usepackage{fancybox}
 \usepackage{authblk}
 \usepackage{siunitx}
 \usepackage{booktabs}
 \usepackage{multirow}
 \usepackage{soul}
 \usepackage{microtype}
 \usepackage[pdftex]{graphicx}
 \usepackage[pdftex,bookmarksnumbered=true,breaklinks=true]{hyperref}
\usepackage{siunitx}

\newcommand{\eqnref}[1]{Eqn.~(\ref{#1})}		
\newcommand{\figref}[1]{Fig.~\ref{#1}}			
\newcommand{\tabref}[1]{Tab.~\ref{#1}}			
\newcommand{\secref}[1]{Section~\ref{#1}}		
\newcommand{\appref}[1]{appendix~\ref{#1}}			
\newcommand{\pa}{\partial}						
\newcommand{\diff}[2]{\frac{\pa #1}{\pa #2}}				
\newcommand{\qbounce}{{\it{q}}{\sc{Bounce}}}				

\newcommand{\ri}{{\rm i}}						
\newcommand{\re}{{\rm e}}						

\renewcommand{\a}{\alpha}

\newcommand{\n}{\nu}

\renewcommand{\t}{\tau}

\renewcommand{\Xi}{\Xi}

\newcommand{\inv}[1]{\frac{1}{#1}}					

\graphicspath{{}{./figures/}}
\renewcommand\journalref[6]{\renewcommand\thejournalref{{\emph{#2}.}}%
                          \renewcommand\theauthors{#1}%
                          \renewcommand\thetitle{#6}%
                          \title{#6}}
 \pacs{ultracold neutrons; Ramsey spectroscopy; quantum bouncer}
 \journalref{Sedmik, Bosina, Achatz, \emph{et al}}{accepted by: J. Phys.:Conf. Ser}{X}{X}{2019}{Proof of Principle for Ramsey-type \\*Gravity Resonance Spectroscopy with qBounce}
\title{Proof of Principle for Ramsey-type \\*Gravity Resonance Spectroscopy with qBounce}
\author{Ren{\'e I.P.} Sedmik$^1${\footnote{Correspondence: rene.sedmik@tuwien.ac.at}}}
\author{Joachim Bosina$^{1,2}$}
\author{Lukas Achatz$^{1}$}
\author{Peter Geltenbort$^2$}
\author{Manuel Hei\ss{}$^{1}$}
\author{Andrey N. Ivanov$^1$ }
\author{Tobias Jenke$^2${\footnote{jenke@ill.fr}}}
\author{Jakob Micko$^{1,2}$}
\author{Mario Pitschmann$^{1}$}
\author{Tobias Rechberger$^{1}$}
\author{Patrick Schmidt$^{1}$}
\author{Martin Thalhammer$^{1,2}$}
\author{Hartmut Abele$^1$\footnote{abele@ati.ac.at}}
\affil{$^1$Technische Universit\"at Wien, Atominstitut, 1020 Vienna, Austria}
\affil{$^2$Institut Laue Langevin, 38042 Grenoble Cedex 9, France}
\hypersetup{pdfauthor={\theauthors},pdftitle={\thetitle}}
\begin{document}
\maketitle
\begin{abstract}
Ultracold neutrons (UCNs) are formidable probes in precision tests of gravity. With their negligible electric charge, dielectric moment, and polarizability they naturally evade some of the problems plaguing gravity experiments with atomic or macroscopic test bodies. Taking advantage of this fact, the \qbounce{} collaboration has developed a technique -- gravity resonance spectroscopy (GRS) -- to study bound quantum states of UCN in the gravity field of the Earth. This technique is used as a high-precision tool to search for hypothetical Non-Newtonian gravity on the micrometer scale. In the present article, we describe the recently commissioned Ramsey-type GRS setup, give an unambiguous proof of principle, and discuss possible measurements that will be performed.
\end{abstract}
%
\section{Introduction}
\label{sec:intro}
%
Gravity resonance spectroscopy (GRS)~\cite{Nesvizhevsky:2005,Abele:2010,Jenke:2011,Jenke:2014,Cronenberg:2018} has been developed as a tool to search for non-Newtonian gravity, combining the virtues of ultracold neutrons (UCNs) -- namely insensitivity to electrostatic and van der Waals interactions -- with spectroscopy, being the method permitting the highest possible sensitivity known. Massive particles bound between the Fermi potential of a horizontal mirror from below and the gravitational potential of the Earth from above form eigenstates of non-equidistant energies. This system -- known as the `quantum bouncer'~\cite{Gibbs:1975}, has been investigated theoretically~\cite{Langhoff:1971,Gea-Banacloche:1999,Rosu:2002} and experimentally~\cite{Abele:2009,Jenke:2009}. Early efforts and the classical limit are reviewed in Ref.~\cite{Abele:2008}. For UCN, the spatial extent of the wavefunctions is of the order of tens of microns, which makes them very well accessible in the experiment. As the eigenenergies, are non-equidistant and in the range of a few peV, it is possible to uniquely address and trigger transitions between them by acoustic-frequency mechanical oscillations. Furthermore, these energies solely depend on three parameters: the neutron mass~$m$, the reduced Planck constant~$\hbar$, and the Earth's acceleration~$g$. Therefore, by measuring transition frequencies between UCN bound states, GRS is highly sensitive to hypothetical deviations from Newtonian gravity at the micron distance scale. We exploited this fact to set tight limits on chameleon~\cite{Jenke:2014,Ivanov:2013,Ivanov:2016a} and symmetron~\cite{Cronenberg:2018,Brax:2018} dark energy as well as axion~\cite{Jenke:2014} interactions.\\*
In recent years, the search for non-Newtonian forces has received much attention. The development of new techniques such as Cs atom interferometry~\cite{Hamilton:2015,Jaffe:2017}, spin-relaxation of $^3$He~\cite{Petukhov:2010}, or the isoelectric Casimir technique~\cite{Chen:2016}, has lead to impressive reductions of the open parameter spaces of most dark sector models. In order to further reduce these limits, and eventually exclude candidate models for dark energy and dark matter, an improvement of the setups in terms of sensitivity is required. In \qbounce{}, such an improvement can be realized by extending the recently utilized Rabi-type setup ~\cite{Cronenberg:2018} to allow for Ramsey spectroscopy~\cite{Ramsey:1949} with quantum states in the gravitational potential of the Earth~\cite{Abele:2010}.
Ramsey spectroscopy is implemented by splitting the central excitation region of a Rabi setup into two parts that are traversed in sequence by the test particles. These two parts are then separated by an additional passive region, where the particle states may evolve either freely or under the influence of external conditions. Thereby, a long interaction time can be realized without the technical burden of a long excitation region. Such a change, while potentially reducing the neutron count rate and increasing the complexity of the setup, allows us to increase the interaction time and reduce the sensitivity to systematic effects, such as the spectral distribution of neutron velocities. Motivated by this perspective, we have developed a new GRS setup implementing a Ramsey-type scheme.\\*
In the present article, we describe this new setup and the underlying measurement scheme in \secref{sec:setup}, discuss the relevant theory in \secref{sec:theory}, and give a first set of data demonstrating the working principle in~\secref{sec:pop}. Finally, we present a brief outlook on possible applications in~\secref{sec:outlook}.
\begin{figure*}[t]
	\centering
	\includegraphics[scale=0.92]{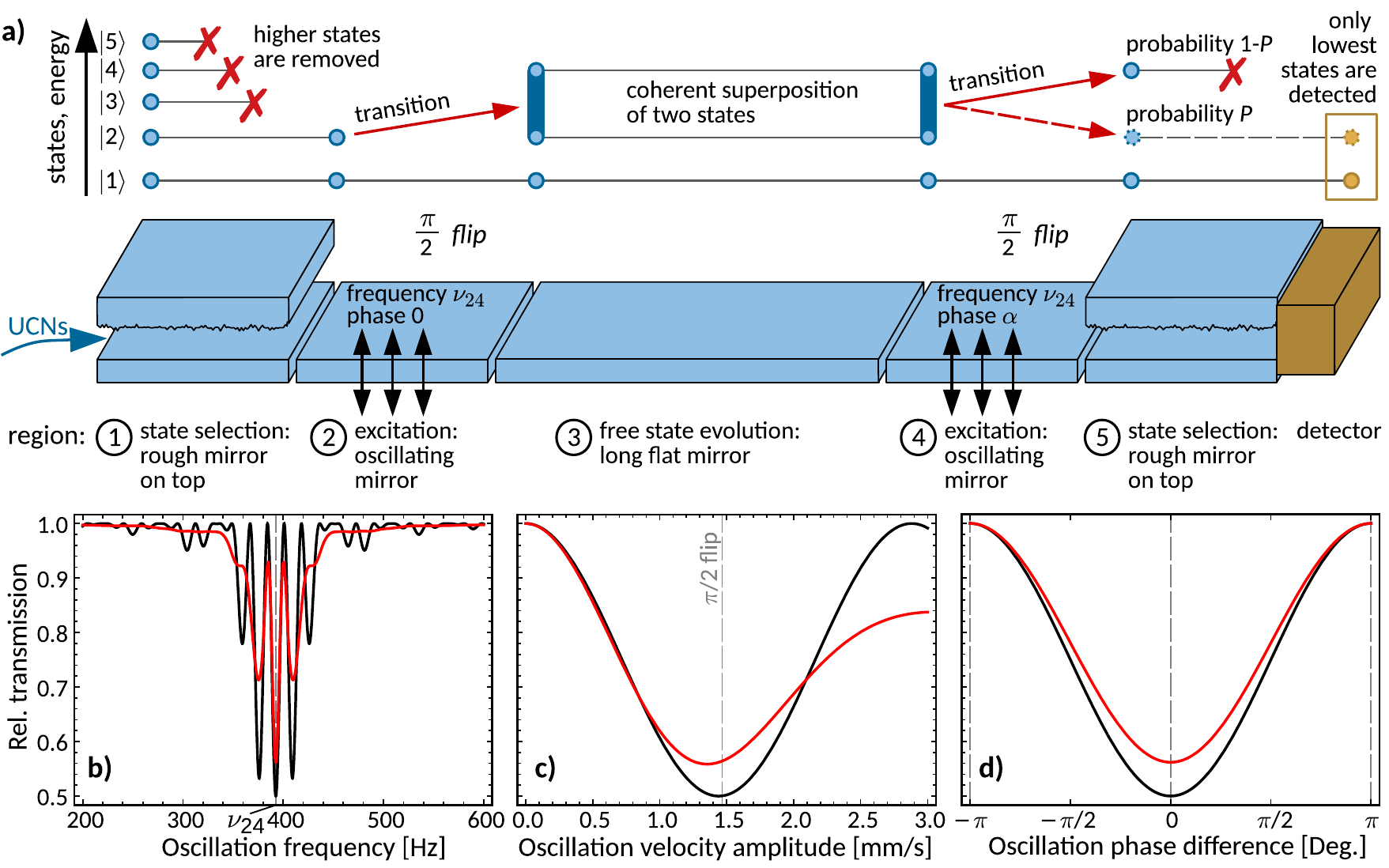}
	\caption{Working principle of Ramsey GRS. a) Illustration of the UCN state evolution in our setup at hand of the transition $|2\rangle\leftrightarrow|4\rangle$. b--d) Theoretical dependence of the relative transmission $1-P$ on frequency, oscillation velocity amplitude $A\omega$, and the relative phase $\a$ between the oscillations in regions 2 and 4.}
	\label{fig:setup_schematic}
\end{figure*}

%
\section{Ramsey GRS -- scheme and setup}
\label{sec:setup}
%
The working principle of GRS relies on the non-equidistant energy levels $E_n$ of UCNs with mass $m$ in a gravitational well. We now illustrate this scheme qualitatively at hand of the transition between the second and the fourth eigenstates, $|2\rangle\leftrightarrow|4\rangle$, in \figref{fig:setup_schematic}.
UCNs enter the setup from the left through a velocity selecting aperture system (not shown). This system consists of two neutron-absorbing blades with vertical opening~$s$, mounted at a distance~$\ell$ and height~$h$  in front of the first region. Region 1 consists of a polished neutron mirror on the bottom with a rough mirror on top, separated by a slit of width~$h_s\ll h,s$. The aperture system selects all neutrons with a horizontal velocity component~$v_x$ fulfilling the condition $\sqrt{g\ell^2/(2h)}\lesssim v_x\lesssim\sqrt{g\ell^2/[2(h+s)]}$ (flight parabola between the openings of the aperture and region 1).\\*
\begin{figure*}[!th]
	\centering
	\includegraphics[scale=0.92]{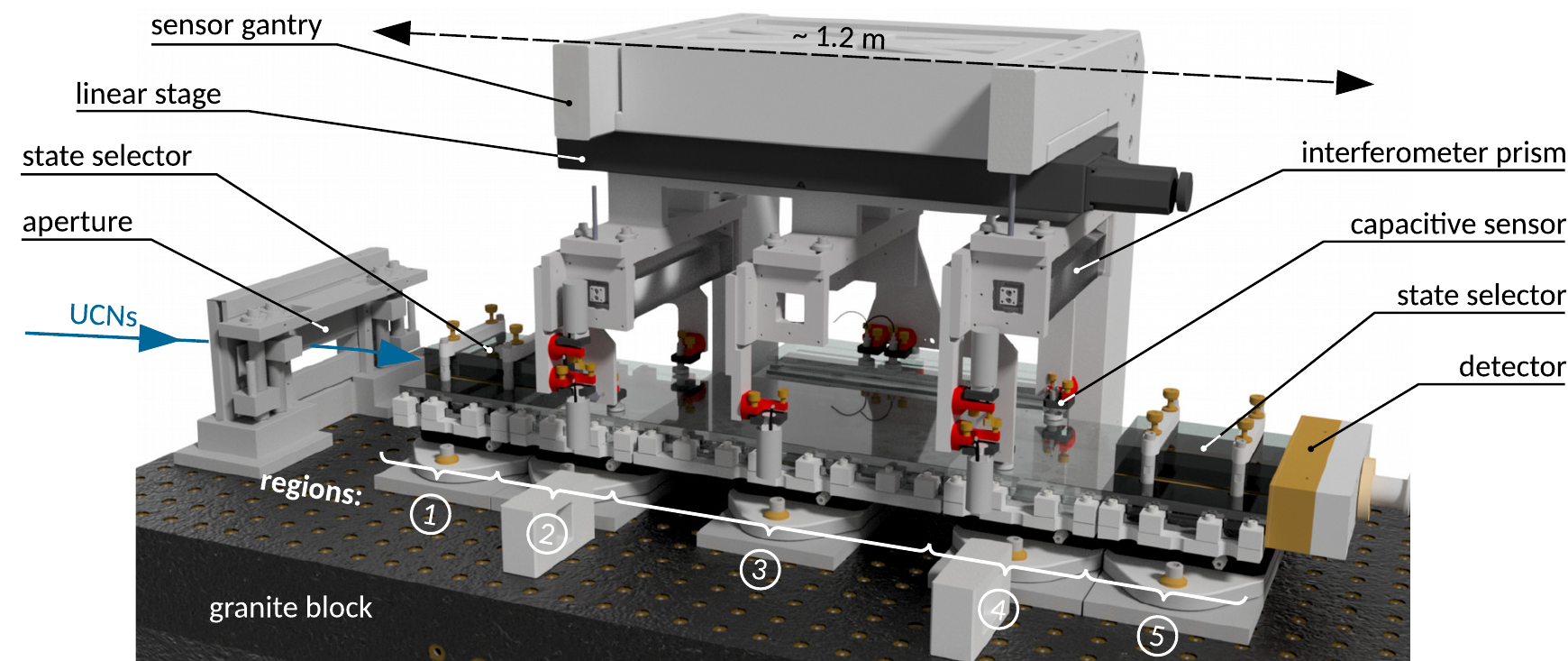} 
	\caption{Simplified view of the Ramsey GRS setup on the granite block.}
	\label{fig:setup}
\end{figure*}
In the first region, we exploit the fact that the spatial probability densitiy of neutrons in higher states extends further upwards than the one of lower states. Therefore, high states interact with high probability with the upper rough surface, whereupon they are scattered out of the system or absorbed~\cite{Nesvizhevsky:2002,Westphal:2007}. Hence, only UCNs in the lowest gravitational states can pass through region 1. The remaining UCNs in region 2 that are in state $|2\rangle$ are excited by mechanical oscillation into a coherent superposition of states $|2\rangle$ and $|4\rangle$. For this excitation we make use of the fact that the resonance frequencies $\omega_{mn}=|E_n-E_m|/\hbar$ are unique, for which we can address transitions between specific states $|n\rangle$ and $|m\rangle$. In region 3
the states are allowed to evolve freely on the horizontal mirror, before a second excitation in region 4 drives the $|2\rangle$ to $|4\rangle$ transition once more (at the same frequency $\nu=\omega/2\pi$ as in region 2). On resonance ($\omega=\omega_{24}$), the state transition into the target state $|4\rangle$ is thereby completed. The state selector in region 5 again selects only the lowest gravitational states.  As all neutrons converted by excitation from $|2\rangle$ to $|4\rangle$ are filtered out in this process, while neutrons in state $|2\rangle$ can pass region 5 with high probability, the rate measurement after region 5 directly reflects the transmission probability $P$, and hence also the transition probability $1-P$. The latter mainly depends on the amplitude $A$, relative phase $\a$, and frequency $\nu$ of the two excitations, and the precise form of the gravitational potential. These dependencies are depicted in the lower part of Fig.~\ref{fig:setup_schematic}  for the $|2\rangle \leftrightarrow |4\rangle$ transition: In (b), the frequency~$\nu$ is swept while keeping $A$ and $\a$ constant, which leads to the well-known frequency pattern of Ramsey spectroscopy. The middle plot (c) shows the dependence of the transmission on the amplitude $\omega A$ for resonant conditions ($\nu = \nu_{24}$ and $\alpha = 0$). Plot (d) on the right side shows a unique feature of Ramsey spectroscopy -- the sinusoidal variation of the transmission with the relative phase $\a$ between the oscillations in region~2 and 4. Here, we have also assumed a resonance condition ($\nu = \nu_{24}$ and $A$ corresponding to a $\pi/2$-flip). In all plots of \figref{fig:setup_schematic}, the black curves are calculated for the mean UCN velocity of \SI{8.8}{\meter/\second}, while the red curves are convolutions of the theory with the (measured) velocity spectrum (see Fig.~\ref{fig:velo_distr}).\\
The split of the excitation in two regions with an intermediate free evolution is the main difference between Rabi and Ramsey spectroscopy. For both methods, the sensitivity increases with the interaction time. In practice, however, the implementation of two smaller excitation regions is technically simpler than one large such region. In addition, Ramsey spectroscopy has the advantage of being less sensitive to a broad horizontal velocity (time of flight) spectrum.

Similar to previous versions, the \qbounce{} Ramsey GRS setup is located at the UCN beam position of the ultracold and very cold neutron facility PF2 at the Institute Laue-Langevin (ILL).
In accordance with the scheme described above, the setup shown in \figref{fig:setup} is comprised of a row of five `regions' (labeled 1--5). Each region consists of a \SI{30}{\milli\metre} thick optically flat BK7 glass mirror coated with a conductive layer of Al and a protective SiO$_2$ layer on top, carried by a 3-axis piezo-electric stage (height, pitch, and roll). These stages have two tasks. The first one is to control the parallel alignment of the mirror's top surfaces to maintain a unified mirror surface guiding the neutrons at the same height without steps from the entry on the left over the five regions into the detector behind region 5. The second task is to create vertical oscillations of the mirrors in regions 2 and 4 to trigger state transitions (see \secref{sec:theory}). For this purpose, an arbitrary function generator (AFG) supplies the stages in region 2 and 4 with independent analog sinusoidal setpoint signals. While both of these signals oscillate at the same frequency $\nu$, the AFG allows to add a phase shift $\a$ between them. 
In regions 1 and 5, additional rough glass plates acting as state selectors are placed at a height $h_s=\SI{22}{\micro\metre}$ above the smooth mirror surfaces\footnote{The separation is set via precision spacers that are clamped between the mirror and the upper glass plate.}. 
The five regions and the detector are rigidly fixed to a massive fine-polished granite block. A tilt sensor with resolution \SI{1}{\micro\radian} is used to drive a feedback circuit controlling three piezo-electric transducers below the granite, thereby leveling the block with respect to the local gravity axis to $\Delta g/g<10^{-10}$. In \figref{fig:setup}, a gantry arches the five regions. Its purpose is to allow an array of capacitive sensors and pentaprisms to sweep over the mirror surfaces to detect deviations in the mirror alignment. PISeca D-510 capacitive sensors are used to detect and correct minor steps between the mirror surfaces to below \SI{1}{\micro\metre} via a feedback circuit controlling the piezo stages (thereby reducing errors in hight, pitch, and roll of the mirrors). The pentaprisms attached to the gantry are part of two independent laser interferometers (not shown, Sios GmbH, SP15000TR and SP2000TR, the latter one being mechanically decoupled from the granite) monitoring the oscillating movement of regions 2 and~4. The entire setup is placed inside a large rectangular vacuum chamber (pressure $\sim\SI{e-5}{\milli\bar}$). Neutrons are supplied from the Steyerl UCN turbine via a series of Ni-coated glass guides separated by thin Al foils and two independent shutter systems. The rate of neutrons provided by the turbine is monitored by a dedicated $^3$He detector attached to the glass guides.\\
Isolation against various disturbances plays an important role in \qbounce{}. Seismic and acoustic vibrations are reduced by the mass (\SI{840}{\kilo\gram}) of the granite block and the vacuum environment. A two-layer Mu-metal shielding inside the chamber achieves an attenuation of external magnetic fields by a factor 100 on the granite surface. Thermally, the environment at the PF2 installation is controlled within $\sim\SI{1}{\kelvin}$ around room temperature.\\
%
\section{Theoretical aspects}
\label{sec:theory}
%
The theory of GRS has been discussed extensively in the literature~\cite{Jenke:2011,Cronenberg:2018}, for which we only give a brief explanation here. UCNs of mass $m$ trapped between the Fermi energy $E_F$ of a vibrating horizontal mirror at $z_m(t)\!=\!A\sin(\omega t\!+\!\phi)$ and the gravitational potential $V_g(z)\!=\!m g z$ of the Earth can be described by the Schr\"odinger equation (with usual conventions for $\hbar$ and $g$)
\begin{align}
 &\ri \hbar\diff{}{t}\psi(z,t)=H\psi(z,t)\,,\label{eq:schroedinger_general}\\
 \text{with } H=&-\frac{\hbar^2}{2m}\diff{^2}{z^2}+m g z +E_F\theta\left[A\sin(\omega t+\phi)-z\right]\,,\nonumber 
\end{align}
and $\theta$ denoting the Heaviside theta function. For $A\to0$, the solutions of \eqnref{eq:schroedinger_general} are known to be the Airy functions $\langle z|n\rangle=\psi_n(z)=\kappa_n {\rm Ai}\left[z/z_0-E_n/E_0\right]$ with normalization constant $\kappa_n$ and the characteristic scales $z_0^3=\hbar^2/2 m^2g$ and $E_0^3=\hbar^2 m g^2/2$~\cite{Langhoff:1971,Abele:2009}. In general, a neutron assumes a mixed state that can be written as a coherent superposition $\psi(z,t)=\sum_n C_{n}(t)\re^{-\ri E_nt/\hbar}\psi_n(z)$, with coefficients $C_n(t)$ that fulfill $\sum_n|C_n(\tau_1)|^2=1$. The corresponding energy levels $E_n=\{1.407,\, 2.459,\, 3.321,\, 4.083, ..\}\,$peV are non-equidistant, which opens the possibility to trigger transitions between specific states $|m\rangle$ and $|n\rangle$ by applying an excitation at circular frequency $\omega_{mn}=2\pi\nu_{mn}=|E_m-E_n|/\hbar$. In this article, we only consider the transition $|2\rangle\leftrightarrow|4\rangle$. Depending on the amplitude $A$ of the oscillation, and the interaction time $\Delta\tau_i=\tau_i-\tau_{i-1}$ of the neutron with the excitation in region $i$, transitions will occur with a certain probability (denoted generically $1-P$ in \figref{fig:setup_schematic} having its maximum at $A\omega\Delta\tau_i\left|\langle m|\partial_z|n\rangle\right|=\pi$ (defining a `$\pi$ flip'). By detecting only states up to $n'$ with $m\leq n'<n$, we only detect those neutrons that do not undergo a transition to higher states. Hence, we can relate the transition probability $1-P$ to the measured transmission rate $r$. As for the UCNs in our experiment the Fermi energy is large compared to the vertical kinetic energy of the neutron, we may assume $E_F\to\infty$, leading to the boundary condition $\psi(z=z_m,t)=0$. The parameter $\phi$ describes an unknown phase (or time) offset between the arrival of the neutron and the zero position of the continuous vibration. For Ramsey spectroscopy, $\phi$ is irrelevant but the difference $\alpha=\phi_4-\phi_2$ between the oscillations in region 4 and 2, respectively, plays an important role, as we will explain further below.

After traversing the first state selector (region 1) neutrons entering region 2 at $t=\tau_1$ from the left (see \figref{fig:setup_schematic}), arrive in exactly one energy eigenstate $E_n$~\cite{Nesvizhevsky:2002,Westphal:2007}. Statistically, these eigenstates have relative probabilities of occurrence $a_n$ with $\sum_n a_n=1$, for which we describe the total state as an incoherent sum $|\psi(z,t)|^2=\sum_{n}a_n|\psi_n(z,t)|^2$. From separate measurements of the $a_n$ and the respective normalized coefficients $b_n$ for the second state selector in region 5~(the method is documented in~\cite{Jenke:2013}), we know that only the first 3 states have significant probabilities of occurrence. For this reason, we focus in the following on the transmission probability of the state $|2\rangle$, which (apart from an offset given by the unaffected states $|1\rangle$ and $|3\rangle$) is the quantity measured in our setup. This effectively reduces the incoherent sum to one single term, $|\psi(z,t)|^2=|\psi_2(z,t)|^2$.

The evolution of $\psi(z,t)$ for $t>\tau_1$ is determined by the evolution of the coefficients $C_{n}(t)$, which is given by
\begin{align}
 \partial_t C_{p}(t)=A\omega\cos(\omega t+\phi)\sum\limits_{n}\re^{\ri\omega_{pn}t}T_{pn}C_{n}(t)\,,\label{eq:coeff_diffeq}
\end{align}
with the transition matrix elements $T_{np}=\langle n|\partial_z|p\rangle=-T_{pn}$.
Analytical solutions of \eqnref{eq:coeff_diffeq} for two and three contributing states are documented in Refs.~\cite{Jenke:2011a,Baessler:2015}. As discussed below, for the $|2\rangle\leftrightarrow|4\rangle$ transition considered in this article, we limit ourselves to the contribution of exactly these two states. Then, the solution of \eqnref{eq:coeff_diffeq} yields a $2\times 2$ coefficient matrix $M(\Delta\tau,\phi)$, such that $\mathbf{C}(\tau_2)=M(\Delta\tau_2,\phi)\mathbf{C}(\tau_1)$, where we use the notation $\mathbf{C}=\{C_{2},C_{4}\}$. We can thus derive the state at $t=\tau_2$ after passing region 2. The same procedure is repeated for region 3, where $M_3\to\mathbb{1}$. In region 4, $M_4(\Delta\tau_4,\phi)$ is again a solution of \eqnref{eq:coeff_diffeq}. Concatenation yields
\begin{align}
 \mathbf{C}(\tau_4)=M_4(\Delta\tau_{4},\phi+\alpha) M_3 M_2(\Delta\tau_{2},\phi)\mathbf{C}(\tau_1)\,,
 \label{eq:transmission_coefficients}
\end{align}
for the state evolution from $\tau_1$ to $\tau_4$. The total transmission probability $P(t)=\int_0^{\infty}\!{\rm d}z\,|\psi(z,t)|^2\approx\int_0^{h_s}\!{\rm d}z\,|\psi(z,t)|^2$ of state $|2\rangle$ through the setup is then obtained from~\footnote{Note that the mixed terms in $|\psi(z,t)|^2$ vanish due to averaging $|\psi(z,t)|^2=\lim_{T\to\infty}(1/T)\int_{-T/2}^{T/2}{\rm d}t_0 |\psi(z,t-t_0)|^2$ over the unknown arrival time $t_0$ of the neutron.}
\begin{align}
 P(\tau_5)\approx\hspace{-3pt}\int\limits_0^{h_s}\!{\rm d}z\,|\psi(z,\tau_5)|^2&=b_2|C_{2}(\tau_4)|^2a_2\,.\label{eq:probability_final}
\end{align}
We would like to stress that the phase $\alpha$ between the excitations in regions 2 and 4 is a free parameter of the experiment that influences $C_2$. The approximation of a two-state system applied here and in Refs.~\cite{Jenke:2011,Jenke:2014,Cronenberg:2018} is convenient to obtain analytic expressions for the $M_i$, but neglects contributions by multiple transitions. Such contributions can be important if several transition frequencies are in close vicinity to each other~\cite{Jenke:2014}. As a measure for this `vicinity' we use the half width of the Rabi fringe (one oscillating region), which can be approximated by the inverse of the interaction time $\Delta\nu_{\rm fringe}\approx1/(2\Delta \tau_2)\approx\SI{29.4}{\hertz}$. In the present case, we investigate the $|2\rangle\leftrightarrow|4\rangle$ transition ($\n_{24}=\SI{392.57}{\hertz}$) with closest neighbours being the transitions $|3\rangle\leftrightarrow|5\rangle$ ($\n_{35}=\SI{352.54}{\hertz}$) and $|1\rangle\leftrightarrow|3\rangle$ ($\nu_{13}=\SI{462.93}{\hertz}$)~\footnote{There can be transitions between higher states $n>5$ that have closer resonance frequencies, such as $|8\rangle\leftrightarrow|11\rangle$ ($\nu_{8\,11}=\SI{390.27}{\hertz}$) or $|7\rangle\leftrightarrow|10\rangle$ ($\nu_{7\,10}=\SI{405.64}{\hertz}$). Such transitions, however, have vanishing small occupation numbers, for which their contribution can be neglected.}. Both of these neighbours are spaced by more than $\Delta\nu_{\rm fringe}$ from $\nu_{24}$. Another way to estimate the `vicinity' is the spectator shift~\cite{Baessler:2015} of $\nu_{24}$ for our setup, which is negligible with $\SI{0.9}{\hertz}$. Using these arguments, we validate the use of a two-state approximation to solve \eqnref{eq:coeff_diffeq} for $C_{2}$ and $C_{4}$. In the computation of $P$, we take into account the spectrum $\eta_v$ of horizontal velocities of the UCNs (see \secref{sec:pop}) by averaging \eqnref{eq:probability_final} over $\eta_v$.

As explained in Sections \ref{sec:setup} and \ref{sec:pop}, in the experiment we determine $P$ from a measurement of the rate $r$ of neutrons transmitted through the setup. While in recent investigations~\cite{Jenke:2014,Cronenberg:2018} of hypothetical non-Newtonian potentials $V(z)$, the transition frequencies between states were determined from $r(\nu)$, in the present article we focus on the dependence $r(\alpha)$, which provides a unique feature of Ramsey spectroscopy. This can be understood as follows. The first `$\pi/2$ flip' in region 2 brings the neutrons into a coherent superposition of the two states. This superposition then freely evolves during the passage of region 3 over a time $\Delta\tau_3$. We could now speculate that the information about the phase $\phi_2$ of the oscillators in region 2 is lost while traversing region 3. In this case, the oscillation in region 4 would represent a statistically independent excitation (in the Markov sense) inducing an independent $\pi/2$ flip. Hence, the initial phases $\phi_2$ and $\phi_4$ entering in the solutions of \eqnref{eq:coeff_diffeq} in regions 2 and 4, respectively, would be independent as well. Then, the total transmission probability would depend on
\begin{align}
|C_{2}(\tau_4)|^2\!=\!\inv{4\pi^2}\hspace{-4pt}\iint\hspace*{-5pt}{\rm d \phi_2}\,{\rm d \phi_4}\left|\left[M_4(\phi_4)M_2(\phi_2)\right]_{11}\!{C_2}(\tau_1)\right|^2\hspace*{-4pt},
\label{eq:2state_2rabi}                                                                                                                                                                                                                                                                                                                                                                                                                                                                                                                                                                                                                                                                                                                                                                                                                                                                                                                              \end{align}
which just represents two Rabi oscillators in series, independent of the relative phase $\phi_4-\phi_2$. If, on the other hand, the phase information is preserved over $\Delta\tau_3$, then we only have one single unknown phase $\phi$, resulting in a single integration of \eqnref{eq:transmission_coefficients} over $\phi\in[0,\,2\pi]$. Here, an experimentally introduced offset $\alpha$ is retained in the final transmission rate. For $\a\to\pi$, the second $\pi/2$ flip inverts the first one, resulting in a revival of the initial state preparation and hence no drop in transmission. Inversely, for $\a\to 0$, we have two synchronized $\pi/2$ flips, resulting in a complete state inversion, and hence the maximum drop in rate.
For $-\pi<\a<\pi$, the transmission rate shows roughly a sinusoidal dependence $-\cos \a$~\cite{Ramsey:1951}. Classically, one would expect that the transmission is independent of the relative phase between mechanical oscillations. We have confirmed this expectation by means of Monte Carlo simulations (see \appref{app:mc}).\\*
Therefore, an experiment in which a sinusoidal dependence of the rate $r$ on the relative phase between the oscillations in regions 2 and 4 is observed, can be considered as an unambiguous demonstration of Ramsey spectroscopy with purely mechanical excitation in the acoustic frequency range.
%
\section{Proof of principle}
\label{sec:pop}
%
In the experiment, we record the rate $r$ of UCNs transmitted through our setup as a function of the frequency, amplitude, and relative phase of the oscillation. In the absence of any oscillation, the transmission rate $r_0=\left.r\right|_{A=0}$ (`zero rate') is defined by the reactor power (neutron flux), the state selectors in regions 1 and 5, loss due to imperfect vertical alignment of our mirrors, loss off the sides of the mirrors, and the detector efficiency. As these factors normally stay constant within $\sim 1$\% over the duration of an experimental cycle, we normalize all results with respect to $r_0$. During the ILL's experimental cycle no. 183 we have recorded 15 data points for $r_0$ and 17 data points with oscillation at frequency $\nu=\SI{392.625}{\hertz}\approx\nu_{24}$ and nominal oscillation amplitude $2\pi A\nu=\SI[per-mode=symbol]{1.47}{\milli\meter\per\second}$, for several different settings of $\alpha$. Over the whole cycle, we observed a constant detector background rate $r_{bg}=0.594\pm0.053\,$mcps (124 neutrons, \SI{58}{\hour}) that is subtracted from all recorded data points prior to normalization, $r_{rel}=(r-r_{bg})/(r_0-r_{bg})$.\\*
Due to software- and electricity problems over a period of several days we were forced to restart the controllers of the piezo-electric stages within our setup, leading to offsets in the neutron mirror positions before and after the error. While we were able to realign the edges of the mirrors to within $\sim$\SI{0.5}{\micro\metre} using capacitive and interferometric measurements, slight residual differences in this alignment have changed $r_0$. We thus discard any data taken during the anomaly, and normalize our results with respect to two different rates, $r_{0,pre}=17.40\pm0.35\,$mcps (9 data points, 2911 neutrons, for data taken before the anomaly), and $r_{0,post}=20.05\pm0.58\,$mcps (6 data points, 1912 neutrons, for data taken after the anomaly). 
The actually applied phase $\a$ was determined from the linearized output of our laser interferometer (Sios SP2000TR). Two of the interferometer's beams were constantly monitoring the surfaces of regions 2 and 4, respectively. The resulting displacement data were Fourier-filtered to a bandwidth of \SI{20}{\hertz} around $\nu_{24}$. We then obtained $\a$ and $A$ from the filtered data by multiplication with normalized sine and cosine functions at frequency $\nu$ and subsequent temporal averaging over the length of one interferometric measurement (\SI{1}{\second}). This procedure represents a software implementation of a four-quadrant lock-in amplifier.\\
\begin{figure}
 \centering
 \includegraphics[width=80mm]{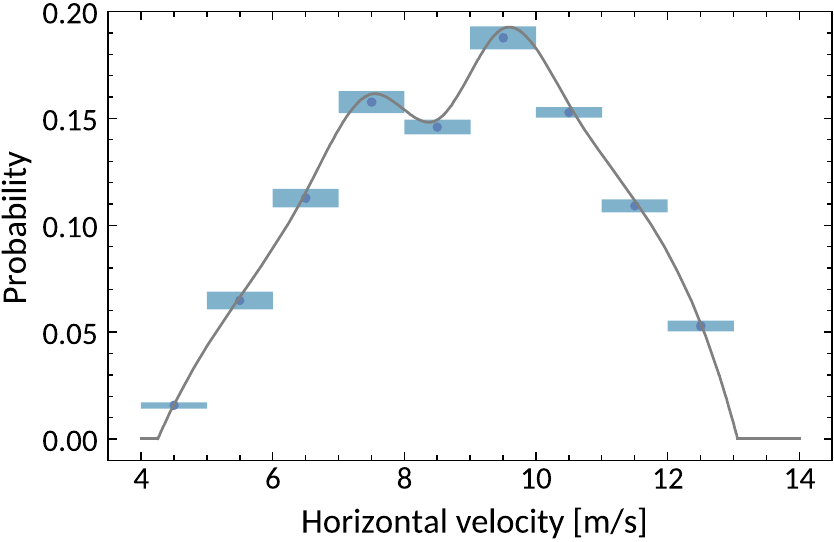}
 \caption{Measured horizontal velocity distribution of neutrons at the exit of the first state selector. The solid line is a spline interpolation intended as a guide to the eye. The boxes represent the selected velocity interval and the uncertainty in the measurement.\label{fig:velo_distr}}
\end{figure}
The sum of the interaction times $\Delta\tau_i$ of the neutrons with the mirrors in regions 2--4 is a critical parameter that defines the width of the transition curves shown in \figref{fig:setup_schematic}b and, thereby, directly scales the achievable precision with which the resonance frequencies can be mesasured. In our experiment, the neutrons are not monochromatic, but posess a certain velocity distribution~$\eta_v$. This distribution is measured by detecting the UCN rate for different settings of the height~$h$ and slit opening~$s$  of the aperture mechanism in front of region~1.
The result is depicted in Fig.~\ref{fig:velo_distr}.
\begin{figure}[!ht]
 \centering
 \includegraphics[width=81mm]{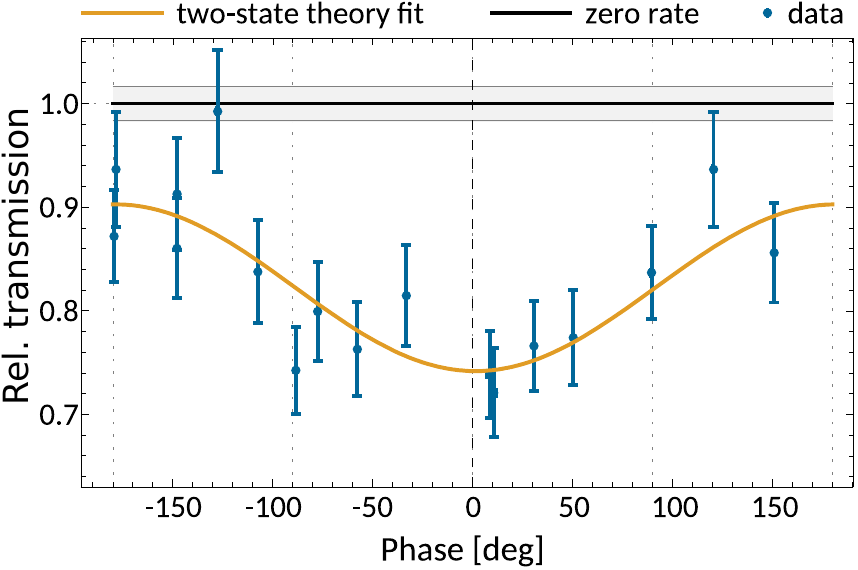}
 \caption{Measured dependence of the relative transmission $r_{rel}$ on the relative angle between the two excitations, together with a theory fit (orange line). A constant (expected classically and if coherence would be lost) is excluded by $>99.99\%$. The grey band shows the average relative zero rate and its uncertainty.\label{fig:sine_fit}}
\end{figure}
\begin{table}[!bh]
\centering\setlength\tabcolsep{4 pt}
 \caption{Results of a least-squares fit of two-state Ramsey theory to the data in \figref{fig:sine_fit}. The fit has 15 degrees of freedom and we compute the fit probability $p$\label{tab:fit_params}.}
 \begin{tabular}{l r@{.}l | l r@{.}l}
 \hline\hline
  \multicolumn{3}{c|}{fit parameters} & \multicolumn{3}{c}{goodness of fit}\\
 \hline
 $r_{off}$&0&$718\pm0.022$ & $\chi^2_{red}$&0&915\rule[10pt]{0pt}{1pt}\\
 $\gamma$ &0&$184\pm0.036$ & $p$&0&55\\
 \hline\hline
 \end{tabular}
\end{table}
Finally, we fit the experimental data on the relative transmission $r_{rel}$ in dependence on $\a$ by a simple model, as described below.
The transmission not only depends on the phase, frequency, and amplitude of the oscillation, but also on the coefficients $a_n$ and $b_n$ determined by the state selectors, and the horizontal velocity distribution $\eta_v$. 
Here, we assume $a_n\approx b_n$, which is valid in our experiment within $\sim 10\%$. Then, from $\sum_n a_n=1$ and $\sum_n |C_{n}(\tau_1)|^2=|C_{2}(\tau_1)|^2=1$ after the first state selector, it follows that the relative amplitude $\gamma$ of the $|2\rangle\leftrightarrow|4\rangle$ Ramsey fringes is mainly determined by the fraction $(a_2+a_4)/\sum_j a_j$ of states at $t=\tau_1$ that participate in the excitation process. As noted in \secref{sec:theory} already, we know that $a_4\approx0$, leaving $a_2$ as single parameter determining the amplitude of the Ramsey fringe. As is illustrated in \figref{fig:setup_schematic}, other states (mainly $|1\rangle$ and $|3\rangle$) will pass through the setup unaffected by the oscillation. These states create a background (offset) in the relative transmission. As argued in \secref{sec:theory}, we can restrict ourselves to a two-state approximation~\cite{Abele:2012}. Hence, only $C_{2}(\t_4)$ and $C_{4}(\t_4)$ are determined by \eqnref{eq:transmission_coefficients}, describing the Ramsey fringe dependence on $\a$, while all other states are ignored. Under these assumptions, $P$ would result in a Ramsey fringe of amplitude $\sim 1$. In order to account for the contributions of other states (again, mainly $|1\rangle$ and $|3\rangle$) as well as losses, we introduce an amplitude parameter $\gamma$, and an offset $r_{off}$ to scale and shift the theoretical curve~\footnote{Losses occuring due to vibration create an offset between the maximum of the fringe and 1. For a perfect state preparation of only neutrons in state $|2\rangle$, the fringe would then extend from this maximum towards zero ($=2\gamma$). However, the sum of transmitted neutrons in states other than $|2\rangle$ reduces the amplitude, for which we adjust $r_{off}$ and $\gamma$ in the fit.}. Eventually, this results in the fit function
\begin{align}
 &P_{\hspace{-0.5pt}fit}(\a,\hspace{-0.5pt}\gamma,\hspace{-0.5pt}r_{off})\hspace{-1pt}=\hspace{-1pt}r_{off}\hspace{-1pt}+\gamma P_{th}(\nu,A,\a,\mathbf{a}\!=\!\{0,\hspace{-0.7pt}1,\hspace{-0.7pt}0,\hspace{-0.7pt}0,\hspace{-0.7pt}...\hspace{-0.7pt}\}),\label{eq:fitfun}\\
 &\text{ with }P_{th}(\nu,A,\a,\mathbf{a})=\hspace{-1.5pt}\inv{2\pi}\hspace{-2pt}\iint\hspace{-5pt}{\rm d}v\,{\rm d}\phi\,\eta_v P(\tau_5(v),\phi,\nu,A,\a,\mathbf{a}),\nonumber
 \end{align}
 and the transmission probability $P$ defined in \secref{sec:theory}.
 
 In order to verify the demonstration of Ramsey GRS, we test two different hypothesis against our data on $r_{rel}(\alpha)$ by performing least squares fits. In this procedure, we weight errors by the experimental standard deviations. First, we fit a constant function $P_c={\rm const}$ with 16 degrees of freedom, resulting in a reduced $\chi^2=2.98$ ($p$-value $5\times10^{-5}$). Hence, a classical explanation (see \appref{app:mc}) and the case that the phase information would be lost between the two excitation regions, that both could be represented by a constant, are excluded by $>99.99$\%. If instead, we fit $r_{rel}(\a)$ to $P_{fit}(\a,\gamma,r_{off})$ as defined in \eqnref{eq:fitfun} (with fixed $\nu$, $\a$, and $A$ from the interferometric measurements, and $\eta_v$ from \figref{fig:velo_distr}), which represents a simplified quantum mechanical treatment of Ramsey spectroscopy, we obtain the results shown in \figref{fig:sine_fit} and \tabref{tab:fit_params}. These results confirm the observation of mechanically driven Ramsey GRS.\\
Note that a slightly less artificial parametrization $P_{alt}=r_{off}+ P_{th}(\nu,A,\a,\mathbf{a}=\{0,a_2,0,0,...\})$ is equivalent and yields exactly the same goodness of fit as the model in \eqnref{eq:fitfun}, with $r_{off}=0.72\pm0.02$, $a_2=0.43\pm0.04$, which is compatible with independent measurements of the spatial probability distribution. It is interesting to note that the theoretical prediction for $P_{fit}$ is very similar to a sine function. A respective fit to $P_{lin}=A+B\sin(\a+\a_{off})$ results in $\chi_{red}^2=1.14$ ($p-$value 0.32). A sine may thus serve as a convenient test function avoiding the technical effort of computing the full theory.\\*[-16pt]\mbox{}
%
\section{Conclusion and Outlook}
\label{sec:outlook}
%
During the past decade, we have developed and extended the \qbounce{} experiment from a UCN quantum bouncing ball measurement of gravitational bound states in the Earth's gravitational potential to gravity resonance spectroscopy (GRS). While with the first Rabi-type version of GRS we have been able to set tight limits on a variety of non-Newtonian dark sector models, the further restriction or exclusion of these models demands for an increase in sensitivity. Following this demand we extended the \qbounce{} GRS setup from a three-region Rabi implementation to a five-region Ramsey-type configuration. The four-fold interaction time (length) directly relates to a proportional increase in sensitivity.\\
In the present paper we give an unambiguous proof of principle for Ramsey GRS. Yet, it remains to reduce systematic effects before reaching the full potential of the new setup. The main limitations at the moment are low counting rates, imperfect state preparation, and alignment of the mirrors. Residual transmission of states $n\gtrsim 2$ limits the achievable contrast of the fringes, and thereby reduces sensitivity.
In future experiments, we aim to achieve contrasts around 70\%, as in previous measurements~\cite{Jenke:2011,Cronenberg:2018}. 
The shape of the zero order interference fringes (near transition frequencies) in Ramsey spectroscopy is far less sensitive to a broad velocity distribution than in Rabi spectroscopy. This insensitivity provides an option to increase the neutron rate rate by opening up the velocity-selecting aperture mechanism. Finally, the alignment and characterization of our mirrors needs to be improved. Other systematic effects, such as tilt of the experiment with respect to gravity and vibrations, only give insignificant contributions and can therefore be neglected for the analysis presented in this article.\\
Besides the aforementioned precision spectroscopy to limit hypothetical dark sector models, Ramsey GRS will be used to test the zero charge of the neutron by adding an electrode above the free interaction region~\cite{Durstberger-Rennhofer:2011}. Another application would be to search for 
spin-dependent energy shifts due to axion or other hypothetical interactions. Such measurements require the addition of a well-controlled magnetic field and spin-selective detection -- both of which we have already demonstrated in a previous version of \qbounce{}~\cite{Jenke:2014}.
\\
On the one hand, the inherently low counting rate of a GRS experiment is a drawback of the present technical realization -- a problem that might be solved in the future. On the other hand, neutrons with their negligible electric charge and electric dipole moment avoid many of the systematic background effects plaguing setups using other test objects, including atoms used in interferometry~\cite{Hamilton:2015}, and macroscopic bodies~\cite{Adelberger:2009,Chen:2016,Sedmik:2018}. Therefore, we consider GRS with UCNs a promising tool to search for physics beyond the standard model; a tool that yields significant potential for further improvement.
%
\section*{Acknowledgments}
We acknowledge financial support from the Austrian Fonds zur F\"orderung der Wissenschaftlichen Forschung (FWF) under contract numbers W1252-N27 (doctoral programme DKPI), I529-N20, 531-N20, W1252, P-26781-N20, and P-31702-N27, and from the German Research Foundation (DFG) as part of the Priority Programme (SPP) 1491 `Precision experiments in particle and astrophysics with cold and ultra-cold neutrons'. We also acknowledge support from the French L'Agence nationale de la recherche (ANR) under contract no. ANR-2011-ISO-007-02, Programme Blanc International--SIMI4-Physique.
\appendix
\section{Classical Monte Carlo simulation}
\label{app:mc}
\begin{figure}[!ht]
 \centering
 \includegraphics{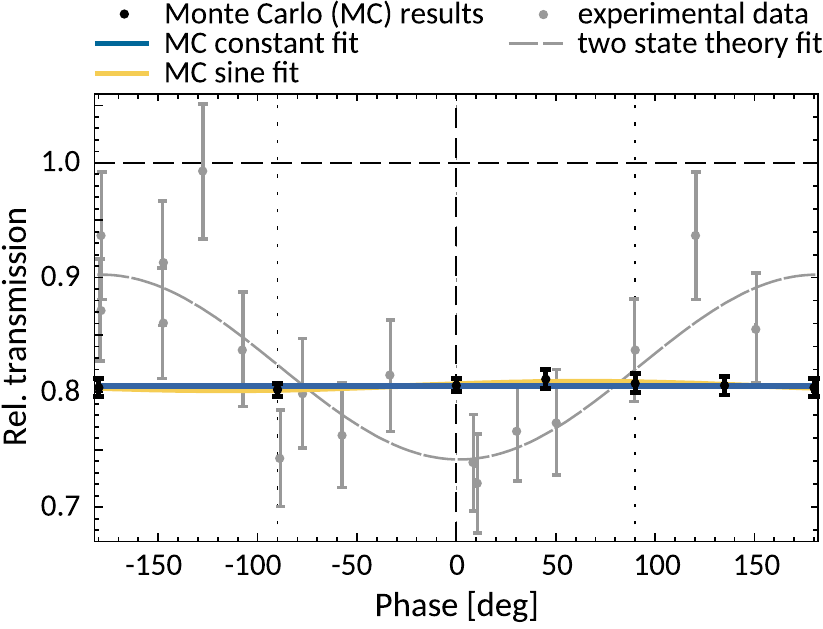}
 \caption{Results of a classical Monte Carlo simulation of the transmission through our setup for the same experimental parameters used in measurements. Experimental data given for comparison are equivalent to those in \figref{fig:sine_fit}.\label{fig:mc}}
\end{figure}

In order to obtain a classical estimate for the expected transmission as a function of the relative phase between the two excitations, we have simulated the neutron's movement as a bouncing ball on a classical trajectory. We approximated the neutrons as point-like massive particles, which reflect elastically on rigid horizontal flat mirror surfaces. Classical parabolic trajectories are then computed analytically between two contacts with mirrors. As starting parameter each neutron receives a horizontal velocity according to the distribution in \figref{fig:velo_distr}, an initial vertical height evenly distributed between 0 and \SI{22}{\micro\metre}, and a vertical velocity evenly distributed within limits that result in a maximal height of \SI{22}{\micro\metre}. The latter two conditions implement the functionality of the state selector in region 1. In addition, the time between the entrance of a neutron and the zero-position of the oscillating mirror in region 2 is chosen to be a random number between 0 and 2$\pi/\omega$. Contacts with the surfaces of regions 2 and 4 change a neutron's vertical velocity upon reflection, depending on the actual mirror position and velocity at time of reflection. Neutrons hitting a side wall of a mirror or exceeding a trajectory height of  \SI{22}{\micro\metre} in region 5 are removed from the simulation. Eventually, the transmission rate is determined from the ratio of surviving neutrons after region 5 to the number of initial neutrons. We have simulated seven different settings for the relative phase between the oscillations in regions 2 and 4, with a minimum of 10000 trajectories per setting. The resulting transmission as a function of phase can be fitted with a constant or a sine function equally well ($p$-values both $>0.95$) but as shown in \figref{fig:mc}, the resulting sine has an amplitude of $(3.8\pm0.8)\times10^{-3}\,$mcps (factor 20 smaller than the quantum mechanical prediction), and a phase offset of $\sim70^\circ$. These results rule out a classical explanation for the observed dependence in \figref{fig:sine_fit}.
\mbox{}\\*[-8pt]\mbox{}

\end{document}